\documentclass[twocolumn,showpacs,nofootinbib,preprintnumbers,amsmath,amssymb,showkeys]{revtex4}

\usepackage{epsfig}
\usepackage{dsfont}
\usepackage{color}
\usepackage{graphicx}
\usepackage{dcolumn}
\usepackage{bm}

\usepackage{amssymb}
\usepackage{amsmath}
\usepackage{amsfonts}
\usepackage{mathrsfs}
\usepackage{amsthm}

\newcommand{\beq}{\begin{equation}}
\newcommand{\eeq}{\end{equation}}
\newcommand{\bea}{\begin{eqnarray}}
\newcommand{\eea}{\end{eqnarray}}
\newcommand{\nn}{\nonumber \\}

\newcommand\eqn[1]{(\ref{#1})}      
\newcommand\Eqn[1]{Eq.~(\ref{#1})}  
\newcommand\Fig[1]{Fig.~\ref{#1}}  

\newcommand{\C}{\mathcal{C}}
\newcommand{\Pn}{{\cal P}}

\newcommand{\bce}{\begin{center}}
\newcommand{\ece}{\end{center}}

\newcommand{\eps}{\varepsilon}
\newcommand{\beps}{\bar\varepsilon}

\newcommand{\bl}{\bar\lambda}
\newcommand{\leff}{\lambda_{\rm eff}}

\newcommand{\la}{\langle}
\newcommand{\ra}{\rangle}

\begin{document}


\title{Scalar field correlator in de Sitter space at next-to-leading\\ order in a $1/N$ expansion}

\author{F. Gautier} 
\affiliation{Physik Department T70, James-Franck-Stra{\ss}e, Technische Universit\"at M\"unchen, 85748 Garching, Germany}
\author{J. Serreau} 
\affiliation{APC, AstroParticule et Cosmologie, Universit\'e Paris Diderot, CNRS/IN2P3, CEA/Irfu, Observatoire de Paris, Sorbonne Paris Cit\'e\\ 10, rue Alice Domon et L\'eonie Duquet, 75205 Paris Cedex 13, France}

\begin{abstract}

 We study the dynamics of light quantum scalar fields in de Sitter space on superhorizon scales. We compute the self-energy of an $O(N)$ symmetric theory at next-to-leading order in a $1/N$ expansion in the regime of superhorizon momenta, and we obtain an exact analytical solution of the corresponding Dyson-Schwinger equations for the two-point correlator. This amounts to resumming the infinite series of nonlocal self-energy insertions, which typically generate spurious infrared and/or secular divergences. The potentially large de Sitter logarithms resum into well-behaved power laws from which we extract the field strength and mass renormalization. The nonperturbative $1/N$ expansion allows us to discuss the case of vanishing and negative tree-level square mass, which both correspond to strongly coupled effective theories in the infrared. 
 
 \end{abstract}

\pacs{11.10.-z, 04.62.+v}
\keywords{Quantum field theory in de Sitter space, Dyson-Schwinger equations, $1/N$ expansion}
                              
\preprint{TUM-HEP-1015/15}
                              
\maketitle

\section{Introduction}
\label{sec:intro}
Understanding the dynamics of quantum fields in curved spacetime is a topic of general interest with important applications to early Universe cosmology or black hole physics. Prominent examples are the primordial density fluctuations during inflation \cite{Mukhanov:1990me} and the Hawking/Unruh radiation from (analog) black holes \cite{Hawking:1974sw,Unruh:1976db,Brout:1995rd}. The case of de Sitter space has attracted particular attention both because of its relevance for inflationary cosmology and because it provides a simple, maximally symmetric example where the nontrivial aspects of the curved geometry come into play. 

The investigation of quantum field dynamics on de Sitter space has many facets. A general class of studies concerns the solutions of the field equations of motion in a fully de Sitter symmetric state. Renormalizability and the equivalence principle select the so-called Chernikov-Tagirov-Bunch-Davies (CTBD) state \cite{Chernikov:1968zm,Bunch:1978yq} out of the class of Allen-Mottola $\alpha$-vacua \cite{Allen:1985ux,Mottola:1984ar,Anderson:2005hi}. Topical questions include the calculation of radiative corrections to the field dynamics \cite{Shore:1979as,Buchbinder1992,Elizalde:1993ee,Prokopec:2002jn,Onemli:2002hr,Brunier:2004sb,Weinberg:2005qc,Tsamis:2005hd,Boyanovsky:2005px,Sloth:2006az,Seery:2007we,vanderMeulen:2007ah,Serreau:2011fu,Prokopec:2011ms,Arai:2011dd,Serreau:2013koa,Nacir:2013xca,Herranen:2013raa,Onemli:2013gya}, the relevance of global coordinate systems as compared to the (expanding) Poincar\'e patch, of interest for inflationary cosmology \cite{Krotov:2010ma,Akhmedov:2012dn,Anderson:2013ila}, and the relation between Lorentzian and Euclidean de Sitter spaces \cite{Rajaraman:2010xd,Beneke:2012kn,Marolf:2010nz,Hollands:2010pr,Higuchi:2010xt,Hollands:2014eia}. Other studies concern other (possibly non de Sitter invariant) states, e.g., in the context of inflationary cosmology \cite{Danielsson:2002kx,Anderson:2005hi,Onemli:2002hr}, or in relation with the question of the quantum stability of de Sitter space \cite{Mottola:1984ar,Mottola:1985qt,Tsamis:1996qq,Polyakov:2009nq,Krotov:2010ma,Marolf:2010nz,Anderson:2013ila,Akhmedov:2013vka}.

An important issue is the understanding of interacting field theories. Not only are these more difficult to tackle than in flat space for obvious technical reasons, but the nontrivial gravitational field leads to specific effects with no flat space analog. Striking examples are the possibility of curvature-induced phase transitions \cite{Shore:1979as,Buchbinder1992,Elizalde:1993ee}, the decay of massive particles into themselves \cite{Bros:2006gs,Jatkar:2011ju}, the generation of a nonvanishing photon mass \cite{Prokopec:2002jn,Prokopec:2003tm,Prokopec:2007ak}, and the radiative restoration of spontaneously broken symmetries \cite{Ratra:1984yq,Mazzitelli:1988ib,Serreau:2011fu,Lazzari:2013boa,Serreau:2013eoa,Guilleux:2015pma}. In the CTBD vacuum, light scalar fields in units of the spacetime curvature exhibit large, gravitationally enhanced quantum fluctuations on superhorizon scales. Notably, this is at the origin of the scale invariance of the (tree-level) power spectrum of primordial density fluctuations in inflationary cosmology. However, such large fluctuations are also responsible for the strong infrared sensitivity of radiative corrections. Perturbation theory typically exhibits infrared and secular (large time) divergences \cite{Brunier:2004sb,Weinberg:2005qc,Tsamis:2005hd,Onemli:2013gya}. The former are generic for massless bosonic field theories \cite{Blaizot:2003tw} while the latter are characteristic of nonequilibrium\footnote{The nonequilibrium nature of (the Poincar\'e patch of) de Sitter space stems for the cosmological expansion in standard comoving coordinates.} systems \cite{Berges:2004vw}. Both types of (spurious) divergences signal a breakdown of the perturbative expansion and need to be resummed. 

Various resummation methods or genuine nonperturbative approaches have been devised to deal with such issues in de Sitter space. The most prominent one is the stochastic effective theory put forward in Refs.~\cite{Starobinsky:1986fx,Starobinsky:1994bd}. This has been shown to correctly capture the nonperturbative dynamics of superhorizon modes at leading order accuracy both in secular logarithms \cite{Tsamis:2005hd,Miao:2006pn,Prokopec:2007ak} and in infrared enhancement factors \cite{Garbrecht:2013coa,Garbrecht:2014dca}. More recent approaches employ various quantum field theoretical tools such as large-$N$ techniques \cite{Mazzitelli:1988ib,Riotto:2008mv,Serreau:2011fu,Serreau:2013psa,Serreau:2013koa}, renormalization group methods \cite{Burgess:2009bs,Kaya:2013bga,Serreau:2013eoa,Guilleux:2015pma}, the Wigner-Weisskopf approach \cite{Boyanovsky:2011xn,Boyanovsky:2012qs}, reduced density matrices \cite{Boyanovsky:2015tba}, and solutions of the Kadanoff-Baym equations (KBEs) \cite{KB,Berges:2004yj}---the nonequilibrium version of the Dyson-Schwinger equations---and their Boltzmann limit \cite{Kitamoto:2010si,Jatkar:2011ju,Garbrecht:2011gu,Akhmedov:2011pj,Akhmedov:2012pa,Akhmedov:2013xka,Akhmedov:2013vka,Youssef:2013by,Serreau:2013psa,Gautier:2013aoa,Rajaraman:2015dta}. 

In Ref.~\cite{Gautier:2013aoa}, we have studied the KBEs for the correlator of an $O(N)$ scalar field theory with quartic self-interactions using the physical momentum representation of de Sitter correlators \cite{Parentani:2012tx,Busch:2012ne,Adamek:2013vw}. We have obtained an exact analytical solution for the two-point correlator in the regime of superhorizon momenta when the self-energy is computed at two-loop order in perturbation theory. The secular divergences of perturbative calculations of the correlator \cite{Brunier:2004sb,Weinberg:2005qc,Tsamis:2005hd,Onemli:2013gya} are a mere artifact of expanding the latter in terms of self-energy insertions at a finite order. Solving the KBEs amounts to resumming the infinite series of such self-energy insertions---and thus the associated large infrared logarithms---which results in a modified power law behavior of the correlator. This is similar to  the generation of an anomalous dimension for critical systems in statistical physics. From the (resumed) propagator, we could compute various quantities of interest, such as the infrared field strength and mass renormalization, at two-loop order. The applicability of such perturbative computations is limited to not too light fields. The cases of fields with either vanishing or negative tree-level square mass suffer from infrared divergences and require further resummations.

In the present article, we generalize this approach to the nonperturbative $1/N$ expansion at next-to-leading order (NLO). The leading-order (LO) approximation resums the infrared divergences of perturbation theory into a self-consistent (local) mass term. This captures interesting nontrivial physics, such as dynamical mass generation and radiative symmetry restoration \cite{Mazzitelli:1988ib,Serreau:2011fu}, or nonperturbative quantum contributions to non-Gaussian correlators \cite{Serreau:2013psa,Serreau:2013koa}. The NLO approximation involves an infinite series of nonlocal multiloop contributions to the self-energy. For infrared momenta, the latter can be summed in a closed form using the results of Ref.~\cite{Serreau:2013psa}. The NLO self-energy has a similar form as the two-loop expression, however, with different infrared power law exponents. This allows us to obtain an exact analytical solution of the corresponding KBEs in the infrared regime using the techniques developed in our previous work \cite{Gautier:2013aoa}.

We thus obtain the complete spacetime structure of the propagator for superhorizon momenta at NLO in the $1/N$ expansion. As in the two-loop case, the infrared logarithms of perturbation theory resum into a superposition of well-behaved momentum power laws, which characterize the decay of the field correlator at large spacetime separations. We compute the leading infrared behavior for deep superhorizon momenta, from which we extract the field strength and mass renormalization. These receive NLO corrections which are nonperturbative functions of the field self-coupling. We study the cases of vanishing and negative tree-level square mass, which both correspond to effectively strongly coupled regimes in the infrared. Finally, we compute the local field variance, which agrees with the result of the stochastic approach at the same approximation order.

The paper is organized as follows. Section~\ref{sec:model} describes the setup and briefly reviews the formulation of KBEs in the $p$-representation \cite{Parentani:2012tx}. We compute the self-energy at NLO in the $1/N$ expansion in Sec.~\ref{sec:self} and present the solution of the corresponding KBEs in Sec.~\ref{sec:schwinger}. The NLO result requires the solution of an appropriate gap equation for the local mass, which is described in Sec.~\ref{sec:gap}. We discuss our analytical solution for the field correlator in various regimes of interest in Sec.~\ref{sec:disc}, and we conclude in Sec.~\ref{sec:conc}. Finally, some technical details and calculations are presented in Appendixes \ref{appsec:self}--\ref{appsec:variance}.

\section{General setting: KBEs in the  $p$-rep\-re\-sen\-ta\-tion}
\label{sec:model}

We consider an $O(N)$ scalar field theory on the expanding Poincar\'e patch of de Sitter space in $D = d+1$ dimensions. In conformal time $-\infty < \eta < 0$ and comoving spatial coordinates $\bold{X}$, the invariant line element reads (we set the Hubble scale $H=1$)
\beq
d s^2 = \eta^{-2}(-d \eta ^2 + d \bold{X} . d \bold{X}).
\label{eq:ds}
\eeq
The classical action is given by 
\beq
 \mathcal{S} = \int_x \left\{\frac{1}{2} \varphi_a \left(\square - m_{\rm dS}^2\right) \varphi_a - \frac{\lambda}{4!N}(\varphi_a\varphi_a)^2\right\},
 \label{eq:action}
 \eeq
with $\int_x = \int d^D x \sqrt{-g(x)}$ the invariant measure---$g(x)$ is the determinant of the metric---and where summation over repeated indices $a=1,\ldots,N$ is understood. Here, $\square$ is the appropriate Laplace-Beltrami operator and $m_{\mathrm{dS}}^2 = m^2 + \xi \mathcal{R}$ includes a possible coupling to the Ricci scalar  $\mathcal{R} = d(d+1)$. We consider a symmetric state such that $\langle\varphi_a\rangle=0$ and the correlator $G$ and the self-energy $\Sigma$ are diagonal, e.g., $G_{ab}=\delta_{ab}G$. In the rest of the paper we assume a de Sitter invariant state given by the free field CTBD  vacuum in the remote past $\eta\to-\infty$.

The covariant inverse propagator is given by 
\beq
 \label{eq:first_2PI}
 G^{-1}(x,x') = G^{-1}_0(x,x') -  \Sigma(x,x'),
\eeq
where 
\beq
\label{eq:bareprop}
 iG_{0}^{-1}(x,x') = \left( \Box_x - m_{\mathrm{dS}}^2 \right) \delta^{(D)}(x,x'),
\eeq
with $\delta^{(D)}(x,x') = \delta^{(D)}{(x-x')}/\sqrt{-g(x)}$.
Extracting a possible local part from the self-energy,\footnote{One may have to include more complicated structures in the local contribution to the self-energy when discussing ultraviolet renormalization. For instance in $D=4$, there appears a term $ \Box_x \delta^{(4)}(x,x^\prime)$ due to field-strength renormalization \cite{Brunier:2004sb}.} one writes
\beq
\Sigma(x,x') = -i \sigma\delta^{(D)}(x,x') +\bar  \Sigma(x,x'),
\label{eq:self_dec}
\eeq
where the local $\sigma$ part is constant for a de Sitter invariant state. We include it in a redefinition of the mass
\beq
 M^2 = m^2_{\mathrm{dS}} + \sigma \label{eq:gap}
\eeq 
and define, accordingly, the  propagator
\beq
\label{eq:massive}
 iG_{M}^{-1}(x,x') = \left( \Box_x - M^2 \right) \delta^{(D)}(x,x'),
\eeq
in terms of which we have
\beq
 \label{eq:first_2PI-M}
 G^{-1}(x,x') = G^{-1}_M(x,x') -  \bar\Sigma(x,x').
\eeq
The aim of the present work is to solve this equation for $G$ when the self-energy \eqn{eq:self_dec}  is computed at NLO in a $1/N$ expansion. The main difficulty concerns the convolution with the nonlocal kernel $\bar\Sigma(x,x')$ when inverting \Eqn{eq:first_2PI-M}. To this aim, we shall exploit the techniques developed in Refs.~\cite{Parentani:2012tx,Serreau:2013psa,Gautier:2013aoa}, which rely on exploiting the de Sitter symmetries in a physical momentum representation, the so-called $p$-representation.

Exploiting the spatial homogeneity and isotropy in comoving coordinates, one writes
 \beq
 G(x,x') = \int \frac{d^d K}{(2\pi)^d} \,\, e^{i\bold{K}\cdot({\bf X}-{\bf X}')} \tilde G(\eta , \eta^\prime , K).
\label{eq:fourier}
\eeq
De Sitter symmetries guarantee that the correlator admit the following scaling form \cite{Parentani:2012tx}
\beq
\label{eq:prepG}
 \tilde G(\eta , \eta^\prime , K)=\frac{(\eta\eta')^{\frac{d-1}{2}}}{K}\hat G(p,p'),
\eeq
where $p = -K\eta$ and $p^\prime = - K \eta^\prime$ are the physical momenta associated with the comoving momentum $K$ at times $\eta$ and $\eta'$ respectively. Similarly, the $p$-representation of the self-energy is
 \beq
 \tilde \Sigma(\eta , \eta^\prime , K)=(\eta\eta')^{\frac{d+3}{2}}K^3\hat \Sigma(p,p').
\eeq

Solving the Schwinger-Dyson equation \eqn{eq:first_2PI-M} for the propagator $G$ for a given self-energy $\bar\Sigma$ in de Sitter space can be viewed as an initial value problem with initial data to be specified in the infinite past $\eta\to-\infty$. This can be formulated by introducing a closed contour in time---the so-called {\it in-in} or Schwinger-Keldysh formalism \cite{Berges:2004yj}---which allows one to conveniently grab together the various components of Green's functions. Exploiting the way time and momentum are tight together by gravitational redshift in de Sitter spacetime, the time evolution can be traded for a momentum evolution, with initial data to be specified at $p\to+\infty$. Introducing a closed contour $\hat\C$ in momentum, the propagator reads \cite{Parentani:2012tx}
\beq
\label{eq:decomp}
 \hat G(p,p')=\hat F(p,p')-\frac{i}{2}{\rm sign}_{\hat\C}(p-p')\,\hat\rho(p,p')
\eeq
where $\hat F$ and $\hat\rho$ denote the $p$-representations of the statistical and spectral two-point functions respectively. Here, the sign function is to be understood along the contour $\hat\C$; see Ref.~\cite{Parentani:2012tx} for details. Notice the symmetry properties $\hat F(p,p')=\hat F(p',p)$, and $\hat \rho(p,p')=-\hat \rho(p',p)$. The (nonlocal) self-energy $\hat\Sigma(p,p')$ admits a similar decomposition.

The Dyson-Schwinger equations expressed in the {\it in-in} formalism are called the Kadanoff-Baym equations \cite{KB,Berges:2004yj}. In the $p$-representation, they read \cite{Parentani:2012tx}
\begin{align}
\hspace{-.15cm}\left[\partial_p^2 + 1 - \frac{\nu^2 - \frac{1}{4}}{p^2}\right]\! \hat F(p,p')&+ \int_{p}^{\infty} \!\!\! d s \,  \hat \Sigma_\rho(p,s) \hat F(s,p')\nn
\label{eq:p_sd_F}
&=\int_{p'}^{\infty} \!\!\! d s \,  \hat \Sigma_F(p,s) \hat \rho(s,p')
\end{align}
and
\beq									
\left[\partial_p^2 + 1 - \frac{\nu^2 - \frac{1}{4}}{p^2}\right] \hat \rho(p,p')	=\int^p_{p'} \! ds \,  \hat \Sigma_\rho(p,s)  \hat \rho(s,p') \label{eq:p_sd_rho},
\eeq
where we introduced (the last equality defines $\varepsilon$)
\beq
\label{eq:eps}
 \nu=\sqrt{\frac{d^2}{4}-M^2}\equiv\frac{d}{2}-\varepsilon.
\eeq
In the following, we consider light fields in units of the spacetime curvature, i.e.,  $\varepsilon\approx M^2/d\ll1$.

As announced, momentum here plays the role of the time-evolution variable. It is remarkable that the KBEs effectively reduce to a $(0+1)$-dimensional dynamical problem in the $p$-representation. In fact, the space dimensionality is completely hidden in the expression of the self-energy through $d$-dimensional loop integrals.
The statistical function $\hat F$ encodes the information about the actual quantum state of the system. Having in mind an adiabatic switch-on of the interactions, the initial data corresponding to the CTBD vacuum are given by $\hat F(p,p^\prime) |_{p=p^\prime \to + \infty} =1/2$, $\partial_p  \hat F(p,p^\prime) |_{p=p^\prime \to + \infty} = 0$, 
$\partial_p \partial_{p^\prime} \hat F(p,p^\prime)|_{p=p^\prime \to + \infty}=1/2$. The nontrivial initial data for the spectral function is determined by the equal-time commutation relations: $\partial_p \hat  \rho(p,p^\prime)|_{p=p^\prime } = -1$. 

\section{The self-energy at NLO}
\label{sec:self}

We now come to the calculation of the self-energy at NLO in a $1/N$ expansion. This has been discussed in the context of nonequilibrium QFT in Minkowski spacetime in Refs.~\cite{Berges:2001fi,Aarts:2002dj}. In de Sitter spacetime the corresponding expressions have been derived in the $p$-representation in Ref. \cite{Parentani:2012tx}. Here, we briefly review the material relevant for our present purposes. 

\begin{figure}[t!]  
\begin{center}
\epsfig{file=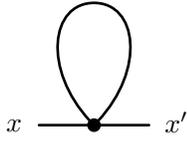,width=2.5cm}
 \caption{\label{fig:Sigma} 
The local contribution to the self-energy $\Sigma(x,x')$; see \Eqn{eq:self_dec}. The internal line in the diagram is given by the full propagator $G(x,x)$.}
\end{center}
\end{figure}

The local contribution in \Eqn{eq:self_dec} can be written in closed form in terms of the exact propagator as
\beq
\label{eq:localsigma}
 \sigma=\frac{\lambda}{6}\left(1+\frac{2}{N}\right)\int_{\bf q}\frac{\hat F(q,q)}{q},
\eeq
where $\int_{\bf q}=\int d^dq/(2\pi)^d$. This is represented diagrammatically in \Fig{fig:Sigma}.
The corresponding expression at NLO is obtained by plugging the (yet unknown) NLO correlator on the right-hand side and discarding terms of ${\cal O}(1/N^2)$. As we shall see below, this actually provides a self-consistent equation for the so-called tadpole mass $M$ defined in \Eqn{eq:gap}. For now, we do not need the expression of $M$ and leave it as a free parameter in the calculation.

\begin{figure}[h!]  
\begin{center}
\epsfig{file=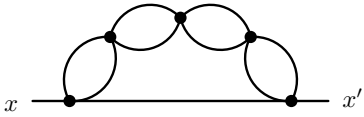,width=5cm}
 \caption{\label{fig:NLO1} 
A typical multiloop diagram contributing to the nonlocal self-energy $\bar\Sigma(x,x')$ at NLO. The latter actually resums all diagrams of similar topology with an arbitrary number $n\ge1$ of bubbles in the upper part, as described by the integral equation \eqn{eq:Np2} in the $p$-representation.}
\end{center}
\end{figure}

The nonlocal part of the self-energy reads, at NLO,
\beq
\label{eq:Np3}
 \hat\Sigma(p,p')=\frac{\lambda}{3N}\,(pp')^{d-3\over2}\!\!\int_{\bf q}\,\frac{r}{q}\,\hat G_M\left(qp,qp'\right)\hat I\left(rp,rp'\right)\,
\eeq
where $r=|{\bf e}+{\bf q}|$, with ${\bf e}$ an arbitrary unit vector, and where the function $\hat I$ resums the infinite series of bubble diagrams represented in \Fig{fig:NLO1}. It solves the following integral equation\footnote{This is directly related to the Dyson-Schwinger equation for the four-point vertex in the large-$N$ limit \cite{Serreau:2013psa}.}:
\beq
\label{eq:Np2}
 \hat I(p,p')=\hat\Pi(p,p')-i\int_{\hat\C} ds \,\hat\Pi(p,s )\hat I(s ,p'),
\eeq
where the momentum convolution on the right-hand side is to be taken on the momentum contour $\hat\C$ to account for the {\it in-in} formulation of the problem and where $\hat \Pi$ is the one-loop bubble integral
\beq
\label{eq:Np1}
 \hat\Pi(p,p')=-\frac{\lambda}{6}\,(pp')^{d-3\over2}\!\!\int_{\bf q}\frac{\hat G_M\left(qp,qp'\right)}{q}\frac{\hat G_M\left(rp,rp'\right)}{r},
\eeq
with $r=|{\bf e}+{\bf q}|$. Notice that we have used the propagator $G_M$ in the NLO expressions \eqn{eq:Np3}--\eqn{eq:Np1}. This is consistent because the LO propagator is given by $G_{M_0}$ with $M_0$ the LO tadpole mass. Here, we shall keep $G_M$ in the intermediate steps of the calculation for notational simplicity and consistently expand the final expressions in $1/N$.

Let us finally mention that $\hat\Pi$, $\hat I$, and $\hat \Sigma$ all have similar decompositions as in \Eqn{eq:decomp} on the momentum contour $\hat\C$. The product $\hat G_M\hat G_M$ in the expression \eqn{eq:Np1} of the function $\hat\Pi$ gives rise to the combinations $\hat F_{\!M}\hat F_{\!M}-{1\over4}\hat\rho_M\hat\rho_M$ for $\Pi_F$ and $2\hat F_{\!M}\hat\rho_M$ for $\Pi_\rho$. Similarly, the product $\hat G_M\hat I$ in \eqn{eq:Np3} yields $\hat F_{\!M}\hat I_F-{1\over4}\hat\rho_M\hat I_\rho$ for $\hat\Sigma_F$ and $\hat F_{\!M}\hat I_\rho+\hat\rho_M\hat I_F$ for $\hat\Sigma_\rho$. The explicit integral equations satisfied by the components $\hat I_F$ and $\hat I_\rho$, obtained from \Eqn{eq:Np2}, read
\beq
\label{eq:Irho}
 \hat I_\rho(p,p')=\hat\Pi_\rho(p,p')+\int^{p'}_{p} ds \,\hat\Pi_\rho(p,s )\hat I_\rho(s ,p')
\eeq
and
\begin{align}
\label{eq:IF}
 \hat I_F(p,p')&=\hat\Pi_F(p,p')-\!\int_{p'}^\infty \!\!ds \,\hat\Pi_F(p,s )\hat I_\rho(s ,p')\nn
 &+\!\int_{p}^\infty \!\!ds \,\hat\Pi_\rho(p,s )\hat I_F(s ,p').
\end{align}

Let us recall the approximation strategy used in Refs.~\cite{Serreau:2013psa,Gautier:2013aoa} for the calculation of correlators for superhorizon momenta $p,p'\lesssim1$. For light fields in units of the spacetime curvature, the relevant dynamics is dominated by the gravitationally amplified superhorizon fluctuations, and we shall neglect the effect of interactions for subhorizon modes. We thus consider a free evolution up to a given scale $\mu\lesssim1$ below which we fully take into account the field self-interaction. In practice, this amounts to restricting the convolution integrals\footnote{The role of subhorizon modes in Eqs.~\eqn{eq:Irho} and \eqn{eq:IF} has been studied in detail in Ref.~\cite{Serreau:2013psa}. They modify the result \eqn{eq:IF_ir} and \eqn{eq:Irho_ir} below by a simple renormalization factor of order unity. } in Eqs.~\eqn{eq:p_sd_F} and \eqn{eq:IF} to $\int_{p}^\infty\to\int_{p}^\mu$ and $\int_{p'}^\infty\to\int_{p'}^\mu$. We obtain a closed set of integro-differential equations which only involve two-point functions for superhorizon momenta. One can show, using the method developed in Ref. \cite{Serreau:2013psa}, that the dominant contribution to the loop integrals in Eqs.~\eqn{eq:Np3} and \eqn{eq:Np1} for $p,p'\lesssim\mu$ comes from loop momenta $qp,qp',rp,rp'\lesssim\mu$. These integrals can thus be evaluated by using the leading infrared behavior of the propagator, that is,
\begin{align}
\hat  F^{\rm IR}_{\!M}(p,p') &= \sqrt{pp'}\, \frac{F_\nu}{(pp')^\nu}, \label{eq:f_loc_ir} \\
 \hat \rho^{\rm IR}_{M}(p,p') &= - \sqrt{pp'}\,{\cal P}_\nu\left(\ln\frac{p}{p'}\right)\label{eq:rho_loc_IR} ,
\end{align}
where $ F_\nu = \left[ 2^\nu \Gamma(\nu)\right]^2/{4\pi}$ and $\Pn_\nu(u) = \sinh(\nu u)/\nu$. The one-loop bubble integral \eqn{eq:Np1} has been computed in  \cite{Serreau:2013psa} and the solution of the integral equation \Eqn{eq:Np2} for infrared momenta has been obtained in closed form. The dominant infrared behaviors of the statistical and spectral components of the infinite series of bubbles $\hat I$ read
\begin{align}
 \hat I_F^{\rm IR}(p,p')&= -\frac{\pi_\rho}{\sqrt{pp'}}  \frac{F_\nu}{(pp')^{\bar \nu-\eps}} \label{eq:IF_ir},\\
 \hat I_\rho^{\rm IR}(p,p')&=\frac{\pi_\rho}{\sqrt{pp'}} \,{\cal P}_{\bar\nu}^\eps\left(\ln\frac{p}{p'}\right), \label{eq:Irho_ir}
\end{align}
where $\eps$ has been defined in \Eqn{eq:eps}, 
\beq
 \pi_\rho=\frac{\lambda F_\nu}{6\eps}\frac{\Omega_d}{(2\pi)^d}\approx\frac{\lambda}{3M^2\Omega_{D+1}},
\eeq
with $\Omega_d=2\pi^{d/2}/\Gamma(d/2)$, $\bar \nu=\sqrt{\nu^2-\pi_\rho}$, and where we defined ${\cal P}_a^b(u)={\cal P}_a(u)e^{-b|u|}$. Equations \eqn{eq:IF_ir} and \eqn{eq:Irho_ir} are valid provided $\pi_\rho\ll1$. More general expressions can be found in \cite{Serreau:2013psa}. Writing $\bar \nu=d/2-\beps$, we thus have $\beps\approx\eps+\pi_\rho/d\ll1$. 

As discussed in Ref.~\cite{Serreau:2013psa}, the function $\hat I$ is of the very same form as the one-loop bubble $\hat\Pi$ with the replacement $\nu\to\bar\nu$. Formally expanding Eqs.~\eqn{eq:IF_ir} and \eqn{eq:Irho_ir} in $\pi_\rho$ generates the infinite series of bubble diagrams, each of which brings an additional power of infrared (secular) logarithms $\pi_\rho\ln (pp')$ and $\pi_\rho\ln (p/p')$. The infinite series of such infrared logarithms resum into the modified power laws \eqn{eq:IF_ir} and \eqn{eq:Irho_ir}. 
 
Using the expressions \eqn{eq:f_loc_ir}--\eqn{eq:Irho_ir}, we can evaluate the dominant infrared behavior of the loop integrals in \eqn{eq:Np3}. This is detailed in Appendix~\ref{appsec:self}. We obtain
\begin{align}
\label{eq:sigmaF}
 \hat \Sigma_F^{\rm IR}(p,p')&=-\frac{\sigma_\rho}{(pp')^{3/2}}  \frac{F_\nu}{(pp')^{\nu-2\gamma}},\\
\label{eq:sigmarho}
 \hat \Sigma_\rho^{\rm IR}(p,p')&=\frac{\sigma_\rho}{(pp')^{3/2}}\,{\cal P}_{\nu}^{2\gamma}\left(\ln\frac{p}{p'}\right),
\end{align}
where we defined $\gamma=(\eps+\beps)/2$ and
\beq
\label{eq:sigma-rho-const}
 \sigma_\rho =\frac{\pi_\rho^2}{N}\,\left(1+\frac{\eps}{2\gamma}\right).
\eeq

We emphasize that, if it is justified to neglect relative corrections of order $\eps$, $\beps$, and $\gamma$ in numerical prefactors, one should not do so in the exponents of the various power law dependences in the momenta. Indeed, such corrections become relevant at large values of $|\ln(p/p')|$ involved in convolution integrals, as discussed in Refs.~\cite{Serreau:2013psa,Gautier:2013aoa}; see also \cite{Garbrecht:2013coa,Garbrecht:2014dca}. 

\section{Solution of the KBEs in the infrared}
\label{sec:schwinger}

It is remarkable that the NLO self-energy, Eqs.~\eqn{eq:sigmaF} and \eqn{eq:sigmarho}, has the very same form\footnote{More precisely, we obtain the same expressions as Eqs.~(19) and (20) in Ref.~\cite{Gautier:2013aoa}, with $s(u)=e^{-(\nu-2\gamma) u}$ and $\sigma(u)={\cal P}_\nu^{2\gamma}(u)$. We check that by setting $\gamma\to\eps$ in the present \Eqn{eq:sigma-rho-const}, we recover the large-$N$ limit of the two-loop expression of $\sigma_\rho$; see Eq.~(21) of Ref.~\cite{Gautier:2013aoa}.} as the two-loop one obtained in \cite{Gautier:2013aoa} up to the  expression of $\sigma_\rho$ and with the replacement $\eps\to\gamma$. This is rooted in the fact that the two-loop self-energy can be written as $\Sigma^{\rm 2-loop}\propto G_M^3\propto \Pi G_M$, with $\Pi$ the one-loop bubble defined in \Eqn{eq:Np1}. Now, the NLO self-energy assumes a similar form with the one-loop bubble replaced by the infinite series of bubble diagrams, that is, $\Pi\to I$; see \Eqn{eq:Np3}. As mentioned above this simply amounts to a modified exponent $\nu\to\bar\nu$. The calculation of the loop integral in \Eqn{eq:Np3} is then essentially the same as in the two-loop case; see Appendix~\ref{appsec:self}.

This observation actually allows us to directly use the results of our previous work \cite{Gautier:2013aoa} for the solution of the KBEs \eqn{eq:p_sd_F} and \eqn{eq:p_sd_rho}. There, we had shown that, for superhorizon momenta $p,p'\lesssim\mu$, these reduce to a single integro-differential equation for a function of one variable, which can be solved exactly by analytical means.  Here, we simply quote the relevant results and refer the reader to Ref.~\cite{Gautier:2013aoa} for details.

The statistical and spectral components of the correlator read, for $p,p'\lesssim\mu$,
\beq
\label{eq:Fmassive}
 {\hat F^{\rm IR}(p,p')}={\sqrt{pp'} F_\nu}\left\{\frac{c_+}{(pp')^{\bar\nu_+-\gamma}}+\frac{c_-}{(pp')^{\bar\nu_--\gamma}}\right\},
\eeq
and 
\beq
 \hat \rho^{\rm IR}(p,p')=-\sqrt{pp'}\, \rho\left(\ln\frac{p}{p'}\right),
\eeq
with
\begin{align}
 \rho (u) &= c_+{\cal P}^\gamma_{\bar\nu_+}(u)+c_-{\cal P}^\gamma_{\bar\nu_-}(u)\nn
 &\approx c_+{\cal P}_{\bar\nu_+-\gamma}(u)+c_-{\cal P}_{\bar\nu_--\gamma}(u),
\label{eq:rho_massive}
\end{align}
where the approximate expression in the second line is valid for $|u|\gtrsim1$. Here, $c_++c_-=1$, with, up to corrections\footnote{The actual solution of the KBEs \cite{Gautier:2013aoa} yields $c_\pm=(\tilde\nu\pm\nu)/(2\tilde\nu)$ and $\bar\nu_{\pm} =\sqrt{\nu^2\pm2\gamma  \tilde\nu+\gamma^2}$, where $\tilde\nu =\sqrt{\nu^2+\sigma_\rho/(4\gamma^2)}$.} ${\cal O}(N^{-2})$,
\beq
  c_-=\frac{\sigma_\rho}{16\nu^2\gamma^2}\approx\frac{\sigma_\rho}{4d^2\gamma^2},
 \eeq
 and
\beq
\label{eq:exponents}
 \bar\nu_\pm=\nu_\pm\pm\frac{\sigma_\rho}{8\nu\nu_\pm\gamma}\approx\nu_\pm\pm\frac{\sigma_\rho}{2d^2\gamma},
\eeq
where $\nu_\pm=\nu\pm\gamma$, and where we neglected terms ${\cal O}(\eps/N,\gamma/N)$ in the final expressions. 

As we already observed in the two-loop approximation \cite{Gautier:2013aoa}, we find that the NLO propagator \eqn{eq:Fmassive}--\eqn{eq:rho_massive} is essentially described by a linear superposition of two free massive field propagators with masses $m_\pm$ defined as
\beq
 \bar\nu_\pm-\gamma=\sqrt{\frac{d^2}{4}-m_\pm^2}\equiv \frac{d}{2}-\eps_\pm.
\eeq
Notice that $m_+<m_-$.
Expanding the above solution in powers of $\sigma_\rho$ formally corresponds to the infinite series of nonlocal self-energy insertions and generates arbitrary powers of infrared logarithms $\sigma_\rho\ln(pp')$ or $\sigma_\rho\ln(p/p')$. These correspond to the infrared secular logarithms obtained in strict perturbative calculations of the field correlator \cite{Brunier:2004sb,Weinberg:2005qc,Tsamis:2005hd,Onemli:2013gya}. Again, the summation of these contributions through the KBEs yields well-behaved infrared power laws with modified exponents $\nu\to\bar\nu_\pm-\gamma$. Going back to spacetime variables, the NLO correlator can be written 
\beq
 G(x,x')=c_+G_{m_+}(z)+c_-G_{m_-}(z),
\eeq
where $G_m(z)$ is the propagator of a free scalar field with mass\footnote{Recall that the present definition of the square mass includes a possible nonminimal coupling to the Ricci scalar.} $m\ll1$ and $z\equiv z(x,x')$ is the de Sitter invariant distance between the points $x$ and $x'$; see Appendix~\ref{appsec:real}.

The solution \eqn{eq:Fmassive}--\eqn{eq:rho_massive} is expressed in terms of the tadpole mass $M$. The latter has to be determined self-consistently from the gap equation \eqn{eq:gap}, with the expression \eqn{eq:localsigma}. 

\section{Determination of the tadpole mass at NLO}
\label{sec:gap}

The tadpole diagram \eqn{eq:localsigma} involves the correlator in the coincidence limit\footnote{The ultraviolet divergence of the latter can be absorbed by a standard renormalization of the parameters of the Lagrangian $m^2$, $\xi$, and $\lambda$ in $D\le4$. The choice of CTBD vacuum guarantees that the required counterterms are the same as in Minkowski space; see, e.g., \cite{Boyanovsky:1997cr,Serreau:2011fu,Serreau:2013psa}. We shall disregard these aspects here and focus on the IR contribution.} $G(x,x)=\la\varphi^2(x)\ra/N$. The latter is dominated by the strongly amplified superhorizon fluctuations and can be evaluated using the expression \eqn{eq:Fmassive}. This yields
\begin{align}
 \frac{\langle\varphi^2(x)\rangle}{N}&=\int\frac{d^dp}{(2\pi)^d}\frac{\hat F(p,p)}{p}\nn
 &\approx\frac{F_\nu\Omega_d}{(2\pi)^d}\left(\frac{c_+}{d-2(\bar\nu_+-\gamma)}+\frac{c_-}{d-2(\bar\nu_--\gamma)}\right)\nn
 \label{eq:variance}
 &\approx\frac{1}{\Omega_{D+1}M^2}\left(1+\frac{1}{N}\frac{\pi_\rho^2}{2d^2\eps\gamma}+{\cal O}\left(N^{-2}\right)\right).
\end{align}
The NLO gap equation \eqn{eq:gap} thus reads
\beq 
\label{eq:gapnlo}
 M^2= m_{\rm dS}^2+\frac{\lambda}{6\Omega_{D+1}M^2}\left[1+\frac{2}{N}\left(1+\frac{(\gamma_0-\eps_0)^2}{\eps_0\gamma_0}\right)\right],
\eeq
where we have used $\pi_\rho/(2d)\approx \gamma-\eps$ and where the subscript $0$ indicates that the corresponding quantity is evaluated at LO. Equation \eqn{eq:gapnlo} is solved as
\beq
 \frac{M^2}{M_0^2}=1+\frac{2}{N}\frac{\leff\left(1+\leff+\leff^2\right)}{(1+\leff)^2},
\eeq
where
\beq
\label{eq:LOmass}
 M_0^2=\frac{m_{\rm dS}^2}{2}+\sqrt{\frac{\left(m_{\rm dS}^2\right)^2}{4}+\frac{\lambda}{6\Omega_{D+1}}}
\eeq
is the LO solution  \cite{Serreau:2011fu} (notice that $M_0^2>0$) and where we introduced the effective coupling
\beq
\label{eq:effcoup}
 \leff=\frac{\lambda}{6\Omega_{D+1}M_0^4}.
\eeq
The relevance of the latter stems from the observation that the effective potential for light scalar fields on superhorizon scales is described by an effective zero-dimensional field theory \cite{Serreau:2013eoa,Guilleux:2015pma}. Equation \eqn{eq:effcoup} corresponds to the dimensionless effective coupling of such a theory at LO in the $1/N$ expansion.

\section{Discussion}
\label{sec:disc}

Let us now discuss some of the peculiar features of the NLO propagator obtained above. In particular, we consider the regime of deep infrared momenta and the corresponding behavior of the field correlator at large spacetime separations as well as the field variance, given by the correlator in the coincident limit. The latter can be compared to the results from other methods such as the stochastic approach or the Euclidean de Sitter approach. We shall examine the relevant quantities in various regimes of parameters from the case of light but massive fields, where perturbation theory applies, to the cases of either zero or negative tree level square mass, described by an effective strongly coupled regime. 
 
\subsection{Deep infrared momenta and large spacetime separations}

The dominant contribution to the regime of deep infrared momenta $p,p'\ll\mu$ and/or large momentum (time) separation $|\ln(p/p')|\propto|t-t'|\gg1$ [where $t=-\ln(-\eta)$ is the cosmological time] is governed by the term with the lowest mass $m_+$ on the right-hand side of either \Eqn{eq:Fmassive} or \eqn{eq:rho_massive}. In this regime, the NLO propagator reduces to that of a renormalized massive field
\begin{align}
 \hat  F^{\rm IR}(p,p')&\approx  \sqrt{pp'}\, \frac{Z\, F_{\nu_{\rm IR}}}{(pp')^{\nu_{\rm IR}}} \label{eq:f_loc_ir_eff}, \\
 \hat \rho^{\rm IR}(p,p') &\approx - \sqrt{pp'}\,Z\,{\cal P}_{\nu_{\rm IR}}\left(\ln\frac{p}{p'}\right)\label{eq:rho_loc_IR_eff} ,
\end{align}
where $Z=c_+=1-c_-$ and, denoting $m_+^2=m_{\rm IR}^2$,
\beq
\label{eq:effmass}
 \nu_{\rm IR}=\sqrt{{d^2\over4}-m_{\rm IR}^2}.
\eeq

The field renormalization factor is completely due to the NLO nonlocal self-energy and reads, in terms of the effective coupling \eqn{eq:effcoup}, 
\beq
\label{eq:IRproperties}
 {Z} =1-\frac{\sigma_\rho}{4d^2\gamma^2}\approx1-\frac{1}{2N}\frac{\leff^2(3+2\leff)}{(1+\leff)^3}.
\eeq
Similarly, the contribution to the infrared mass due to nonlocal self-energy insertions is
\beq
 \frac{ m_{\rm IR}^2}{M^2}=1-\frac{\sigma_\rho}{2d^2\eps\gamma}\approx1-\frac{1}{N}\frac{\leff^2(3+2\leff)}{(1+\leff)^2}.
\eeq
Taking into account the NLO tadpole contribution, we finally obtain 
\beq 
\label{eq:mircouleur}
 \frac{m_{\rm IR}^2}{M_0^2}=1+\frac{1}{N}\frac{\leff\,(2-\leff)}{(1+\leff)^2}.
\eeq
We see that the factor $Z<1$ controls the amplitude of the deep infrared fluctuations in momentum space, whereas $m_{\rm IR}^2$ controls the momentum dependence of the correlator and, in turn, its behavior at large spacetime separation, as described in Appendix~\ref{appsec:real}. 

\subsection{Field variance}

Another quantity of interest is the field correlator in the coincidence limit, which measures the local field variance, $G(x,x)=\la\varphi^2(x)\ra/N$. The latter can be characterized in terms of an effective mass $m_{\rm dyn}$, defined as \cite{Beneke:2012kn,Garbrecht:2011gu}
\beq
\label{eq:def}
 \frac{\langle\varphi^2(x)\rangle}{N}\equiv\frac{1}{\Omega_{D+1}m_{\rm dyn}^2},
\eeq
in analogy with the expression for a free massive field. In fact, such a square mass is nothing but the curvature of the effective potential of the dimensionally reduced theory \cite{Guilleux:2015pma}. We have already evaluated the field variance in \Eqn{eq:variance}. This yields, for the contribution of nonlocal self-energy insertions,
\beq
\label{eq:mdyn}
 \frac{m_{\rm dyn}^2}{M^2}=1-\frac{2}{N}\frac{\leff^2}{1+\leff}.
\eeq
Including the contribution from the NLO tadpole mass, we get
\beq
\label{eq:mdyn0}
 \frac{m_{\rm dyn}^2}{M_0^2}=1+\frac{2}{N}\frac{\leff}{(1+\leff)^2}.
\eeq

This expression can be compared to the result of the stochastic approach, where the local field fluctuations can be computed at all orders of perturbation theory with leading infrared logarithmic accuracy \cite{Tsamis:2005hd}. For light fields, this is equivalently given by the variance of the zero mode on Euclidean de Sitter space \cite{Beneke:2012kn}. We present the calculation of the field variance at NLO in the $1/N$ expansion in the stochastic and Euclidean de Sitter approaches in Appendix~\ref{appsec:variance}, which completely agree with \Eqn{eq:mdyn0}.

\begin{figure}[t!]  
\begin{center}
\epsfig{file=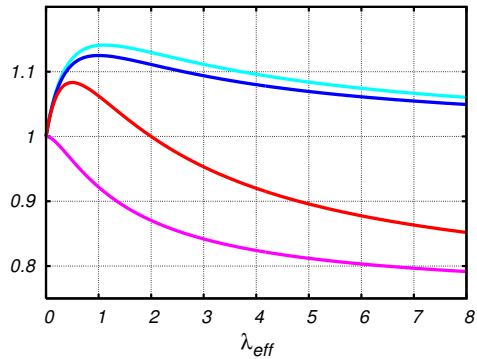,width=7cm}
 \caption{The dimensionless quantities (from bottom to top) $Z$, $m_{\rm IR}^2/M_0^2$, $m_{\rm dyn}^2/M_0^2$, and  $m_{\rm IR}^2/(ZM_0^2)$ as functions of the dimensionless effective coupling $\leff$ for $N=4$. The two lower curves asymptote to $1-1/N$, whereas the two upper ones asymptote to $1$. Deviations from unity are due to both local and nonlocal $1/N$ corrections. \label{fig:ratios}}
\end{center}
\end{figure}

\begin{figure}[t!]  
\begin{center}
\epsfig{file=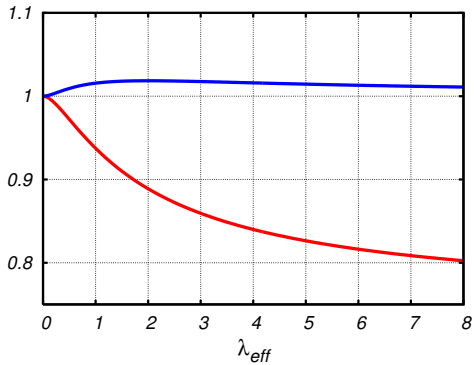,width=7cm}
 \caption{The dimensionless ratios of physical quantities $m_{\rm IR}^2/m_{\rm dyn}^2$ (lower curve) and $m_{\rm IR}^2/(Zm_{\rm dyn}^2)$ (upper curve) as functions of the dimensionless effective coupling $\leff$ for $N=4$. These asymptote to $1-1/N$ and $1$, respectively at large coupling. Deviations from unity are entirely due to nonlocal $1/N$ corrections.\label{fig:ratiosphys}}
\end{center}
\end{figure}

We emphasize that the result \eqn{eq:mdyn0} differs from what one would obtain by evaluating the field variance only from the deep IR behavior, \Eqn{eq:f_loc_ir_eff}. Indeed, this would lead to $\la\varphi^2(x)\ra/N\to Z/(\Omega_{D+1}m_{\rm IR}^2)$. The term $\propto c_-$ in \Eqn{eq:variance} is needed to get the correct variance. However, we mention that the ratio $m_{\rm IR}^2/Z$ turns out to be numerically close to $m_{\rm dyn}^2$ for arbitrary $\leff$. The dimensionless quantities $Z$, $m_{\rm IR}^2/M_0^2$, $m_{\rm IR}^2/(ZM_0^2)$, and $m_{\rm dyn}^2/M_0^2$ are plotted against the effective coupling $\leff$ in \Fig{fig:ratios}. Notice the hierarchy $m_{\rm IR}^2<m_{\rm dyn}^2< m_{\rm IR}^2/ Z$. It is important to notice that all these square masses are equal to $M_0^2$ at LO and that their differences are entirely due to the nonlocal self-energy. Since neither the LO nor the NLO  tadpole square masses $M_0^2$ and $M^2$ are observable, it is interesting to consider ratios of physical quantities such as
\begin{align}
\label{eq:ratio1}
 \frac{m_{\rm IR}^2}{m_{\rm dyn}^2}&=1-\frac{1}{N}\frac{\leff^2}{(1+\leff)^2},\\
\label{eq:ratio2}
 \frac{m_{\rm IR}^2}{Zm_{\rm dyn}^2}&=1+\frac{1}{2N}\frac{\leff^2}{(1+\leff)^3},
\end{align}
whose deviation from unity is entirely an effect of resuming the large infrared logarithms from nonlocal self-energy insertions. The ratios \eqn{eq:ratio1} and \eqn{eq:ratio2} are plotted against $\leff$ in \Fig{fig:ratiosphys}. 

Finally, we note that all the NLO corrections computed in this sections, Eqs.~\eqn{eq:IRproperties}, \eqn{eq:mircouleur}, \eqn{eq:mdyn0}, \eqn{eq:ratio1}, and \eqn{eq:ratio2}, exhibit a nontrivial dependence in the (effective) coupling and are actually bounded functions of the latter due to nonperturbative $1/N$ corrections.  The NLO results thus remain valid for arbitrarily large effective couplings. We shall now discuss the various cases of interest, from the weak coupling, perturbative regime to the strongly coupled one.

\subsection{Perturbative regime} 

Perturbation theory makes sense whenever the quadratic part of the action \eqn{eq:action} gives the dominant contribution to physical observables. This corresponds to the case of light but massive field, with $\sqrt{\lambda}\ll m_{\rm dS}^2\ll1$ \cite{Burgess:2009bs,Garbrecht:2013coa,Garbrecht:2014dca}. In that case, one has $M_0^2=m_{\rm dS}^2[1+\bl_0-\bl_0^2+{\cal O}(\bl_0^3)]$, $\leff=\bl_0[1-2\bl_0+{\cal O}(\bl_0^2)]\ll1$, where we defined 
\beq
 \bl_0=\frac{\lambda}{6\Omega_{D+1}\left(m_{\rm dS}^2\right)^2}.
 \eeq
We get
\beq
 Z=1-\frac{3\bl_0^2}{2N}
\eeq
and
\begin{align}
 m_{\rm IR}^2&=m_{\rm dS}^2\!\left[1+\!\left(1+\frac{2}{N}\right)\!\bl_0-\!\left(1+\frac{7}{N}\right)\!\bl_0^2+{\cal O}\!\left(\bl_0^3\right)\right],\\
 m_{\rm dyn}^2&=m_{\rm dS}^2\!\left[1+\!\left(1+\frac{2}{N}\right)\!\bl_0-\!\left(1+\frac{6}{N}\right)\!\bl_0^2+{\cal O}\!\left(\bl_0^3\right)\right]\!.
\end{align}
These expressions agree with the two-loop results of Ref.~\cite{Gautier:2013aoa} at the relevant order in $1/N$, as they should.\footnote{The effective coupling introduced in Ref.~\cite{Gautier:2013aoa} is $\bar\lambda=\bl_0(1+2/N)$.}
 
\begin{figure}[t!]  
\begin{center}
\epsfig{file=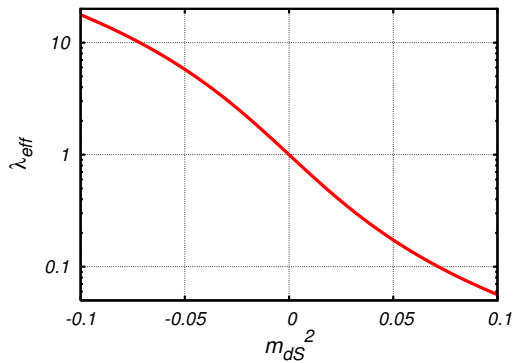,width=7cm}
 \caption{The dimensionless effective coupling \eqn{eq:effcoup} as a function of the tree-level square mass $m_{\rm dS}^2$ for $\lambda=10^{-1}$ in $D=3+1$ dimensions. Large positive values of $m_{\rm dS}^2$ in units of $\sqrt{\lambda/(6\Omega_{D+1})}\approx 0.025$ correspond to perturbatively small effective couplings, whereas small or large negative values correspond to $\leff\approx1$ and $\leff\gg1$, respectively. \label{fig:leff}}
\end{center}
\end{figure}

\subsection{Massless case}

More interesting for our present purposes are theories with very small tree-level mass, such that $\bl_0\gg1$, or with negative tree-level mass, $m_{\rm dS}^2<0$. Both cases correspond to strongly coupled effective theories in the infrared, as measured by the effective coupling \eqn{eq:effcoup}; see \Fig{fig:leff}. We first consider the light mass case. It is well known that, even for massless fields, the scalar self-interactions generate a nonvanishing effective mass. At LO the latter is $M_0^2=\sqrt{\lambda/(6\Omega_{D+1})}$ and the effective coupling $\leff\approx1$. We thus get, at NLO,
\beq
 Z=1-\frac{5}{16N}
\eeq
and
\begin{align}
 m_{\rm IR}^2&=\sqrt{\frac{\lambda}{6\Omega_{D+1}}}\left(1+\frac{1}{4N}\right),\\
 m_{\rm dyn}^2&=\sqrt{\frac{\lambda}{6\Omega_{D+1}}}\left(1+\frac{1}{2N}\right).
\end{align}

\subsection{Negative square mass case}

Finally, in the case of a negative tree-level square mass, the apparently broken symmetry gets radiatively restored by the strong, gravitationally amplified superhorizon fluctuations \cite{Ratra:1984yq,Mazzitelli:1988ib,Serreau:2011fu}. For $\bl_0\ll1$, one has $M_0^2\approx \lambda/(6\Omega_{D+1}|m_{\rm dS}^2|)=\bl_0|m_{\rm dS}^2|$ and the effective infrared theory is strongly coupled: $\leff\approx1/\bl_0\gg1$. We obtain
\beq
 Z=1-\frac{1}{N}
\eeq
and 
\begin{align}
 m_{\rm IR}^2&=\bl_0|m_{\rm dS}^2|\left(1-\frac{1}{N}\right),\\
 m_{\rm dyn}^2&=\bl_0|m_{\rm dS}^2|\left(1+\frac{2}{N}\bl_0\right).
\end{align}

\section{Summary and perspectives}
\label{sec:conc}

Quantum field theoretical calculations in curved spacetimes are technically involved as compared to their Minkowski counterparts because of the nontrivial symmetries of the background geometry. Even the calculation of the basic two-point correlator of a self-interacting scalar field in the maximally symmetric de Sitter space is a notoriously difficult task due to the nontrivial convolution/memory integrals involving the nonlocal self-energy. For cosmological spacetimes, the problem can be cast in the form of a nonequilibrium, initial value setup \cite{Tranberg:2008ae}. A strict perturbative expansion in powers of self-energy insertions leads to spurious secular divergences typical of nonequilibrium systems \cite{Berges:2004vw}. Solving the integro-differential KBEs is formally equivalent to resumming the infinite series of self-energy insertions and yields a well-defined late-time behavior. Together with Refs.~\cite{Gautier:2013aoa}, the present work provides techniques to formulate and solve KBEs in de Sitter spacetime in the case of light fields. Related work for massive fields or for non de Sitter invariant states can be found in Refs~\cite{Kitamoto:2010si,Jatkar:2011ju,Garbrecht:2011gu,Akhmedov:2011pj,Akhmedov:2012pa,Akhmedov:2013xka,Akhmedov:2013vka,Youssef:2013by}.

In the present article, we have computed the two-point correlator of an $O(N)$ scalar field in the regime of superhorizon momenta at NLO in a $1/N$ expansion. Remarkably the solution can be obtained in closed analytical form. This resums the spurious infrared and secular divergences of perturbation theory into a well-defined expression. The nonperturbative $1/N$ expansion allows us to compute various quantities of interest, such as the infrared field strength and mass renormalization, or the local field variance, for arbitrary values of the field self-coupling. In particular, this is a valid approximation scheme for the cases of either vanishing or negative tree-level square mass, which correspond to effectively strongly coupled regimes not accessible by perturbative means. 

The local field variance $\langle\varphi^2(x)\rangle$ obtained here fully agrees with the result of either the stochastic, or the Euclidean de Sitter approaches---which are in fact equivalent for what concerns this observable \cite{Rajaraman:2010xd,Beneke:2012kn}--- as well as the recently proposed nonperturbative renormalization group approach of Refs.~\cite{Serreau:2013eoa,Guilleux:2015pma}. Our result further gives the full spacetime structure of the correlator for noncoincident points at large spacetime separations. To the best of our knowledge, this has never been computed before. It would be of definite interest to perform such a calculation in, say, the stochastic effective theory for a comparison. More generally, it would be interesting to make a precise link between the present approach and the stochastic theory, e.g., along the lines of Refs.~\cite{Garbrecht:2013coa,Garbrecht:2014dca,Gautier:2012vh}.  

Interesting extensions of the present work include the generalization of the techniques developed here to other types of  (e.g., fermionic or gauge) fields and/or to theories with more complicated (e.g., derivative) interactions. Also of interest is the study of other spacetime geometries, such as quasi-de Sitter space and general cosmological spacetimes, or black hole geometries, of interest for the question of radiative corrections to the Hawking radiation \cite{Akhmedov:2015xwa}. 

\appendix

\section{Self-energy at NLO}
\label{appsec:self}

Here, we compute the nonlocal contribution \eqn{eq:Np3} to the NLO self-energy. As explained before, the multiloop nature of the NLO approximation can be captured by the function $\hat I(p,p')$. The latter resums the infinite series of bubble diagrams and eventually takes a similar form as a propagator with modified infrared exponent; see Eqs.~\eqn{eq:IF_ir} and \eqn{eq:Irho_ir} [we refer the reader to Ref.~\cite{Serreau:2013psa} for details concerning the calculation of the functions $\hat I_F(p,p')$ and $\hat I_\rho(p,p')$]. The remaining momentum (loop) integral in \Eqn{eq:Np3} can thus be treated using the methods developed in Ref.~\cite{Serreau:2013psa}. For external momenta $p,p' \lesssim \mu$, it is dominated by internal (loop) momenta such that $qp, qp'\lesssim \mu$, which imply $rp,rp'\lesssim\mu$. We can effectively limit the momentum integration to a sphere of radius $\min(\mu/p,\mu/p')$ and use the leading infrared behaviors \eqn{eq:f_loc_ir}--\eqn{eq:Irho_ir} of the integrands.

We first evaluate the statistical part of the self-energy $\hat \Sigma_F$. In this case the integrand is proportional to $\hat F_{\!M}^{\rm IR} \hat I_F^{\rm IR} - \rho_M^{\rm IR}\hat I_\rho^{\rm IR}/4 \approx \hat F_{\!M}^{\rm IR} \hat I_F^{\rm IR}$. Here, the neglect of the $\rho$-components is justified by the infrared power laws in Eqs.~\eqn{eq:f_loc_ir} and \eqn{eq:IF_ir}, which results from the amplification of infrared fluctuations. The $\rho$-components are not amplified.\footnote{This is a usual feature of the strong (classical) field regime \cite{Berges:2004yj}.} We thus get
\beq
\hat \Sigma_F^{IR}(p,p^\prime) \approx - \frac{\lambda F_\nu^2\pi_\rho }{3N (pp^\prime)^{3/2+\beta}} \int_{{\bf q}}\frac{1}{q^{2\nu} r^{2(\bar\nu - \eps)}} ,
\eeq
where $\beta = \nu - 2\gamma=\bar\nu-2\eps$ and $r=|{\bf e}+{\bf q}|$, with ${\bf e}$ an arbitrary unit vector. The remaining integral is rapidly convergent such that one can safely set the upper bound to infinity. It can then be evaluated, e.g., using the method of Feynman parameters,
\begin{align}
\int_{\bold{q}}\frac{1}{q^{2\nu} r^{2(\bar\nu - \eps)}} &= \frac{\Omega_d}{2(2\pi)^d}\frac{\Gamma(\eps) \Gamma(2\gamma)}{\Gamma(\eps + 2\gamma)}\frac{\Gamma(\nu +\eps)\Gamma(\bar\nu-2\eps)}{\Gamma(\nu)\Gamma(\bar\nu -\eps)}\nn
 &\approx \frac{\Omega_d}{(2\pi)^d} \left(\frac{1}{2\eps}+\frac{1}{4\gamma}\right),
\end{align}
where we have neglected relative corrections of ${\cal O}(\eps,\gamma)$ in the last line. We finally get
\beq
\label{appeq:sigma_F}
\hat \Sigma_F^{IR}(p,p^\prime) = -\frac{\sigma_\rho F_\nu }{(pp^\prime)^{3/2+\beta}},
\eeq
where $\sigma_\rho$ is defined in \Eqn{eq:sigma-rho-const}.  

The integrand for the spectral component $\hat \Sigma_\rho$ involves the combination $\hat F_{\!M}^{\rm IR} \hat I_\rho^{\rm IR} + \rho_M^{\rm IR} \hat I_F^{\rm IR}$. For $p<p'$, it reads
\begin{align}
&\hat \Sigma_\rho^{IR}(p,p^\prime) = \frac{\lambda F_\nu \pi_\rho}{3N}(pp^\prime)^{\frac{d-3}{2}}\nn
& \times \int_{|\bold{q}|\le\frac{\mu}{p'}}  \!\left\{\frac{{\cal P}_{\bar \nu}\!\left(\ln\frac{p}{p'}\right)}{(pp^\prime)^\nu} \!\left(\frac{p}{p^\prime}\right)^{\eps }  \!\frac{1}{q^{2\nu}}+\frac{{\cal P}_\nu\!\left(\ln\frac{p}{p'}\right)}{(pp^\prime)^{\bar\nu-\eps}}  \frac{1}{q^{2(\bar\nu-\eps)}}   \right\}\!.
\end{align}
The remaining integrals are trivially performed,
\beq
 \int_{|\bold{q}|<\mu/p^\prime} \frac{1}{q^{2\alpha}} = \frac{\Omega_d}{(2\pi)^d (d-2\alpha)}\left(\frac{\mu}{p^\prime}\right)^{d-2\alpha} .
\eeq
Repeating the same calculation for $p>p'$, we finally obtain 
\begin{align}
\hat \Sigma_\rho^{IR}(p,p^\prime) &\!=\! \frac{\lambda}{3N}\frac{ F_\nu \pi_\rho}{(pp^\prime)^{3\over2}}\frac{\Omega_d}{(2\pi)^d} \!\!\left\{ \!\frac{{\cal P}_{\bar \nu}^{2\eps}\!\left(\ln\frac{p}{p'}\right)}{2\eps}\!+\! \frac{{\cal P}_\nu^{2\gamma}\!\left(\ln\frac{p}{p'}\right)}{4\gamma} \!\right\}  \nn
&\cong\frac{\sigma_\rho}{(pp^\prime)^{3/2}}{\cal P}_\nu^{2\eta}\left(\ln\frac{p}{p'}\right), \label{appeq:sigma_rho}
\end{align}
where we have neglected relative corrections of ${\cal O}(\eps,\gamma)$ in the constant factors\footnote{As mentioned in the text, it is justified to neglect such corrections in numerical factors [for instance, $\mu^\eps=1+{\cal O}(\eps)$], but not in the momentum dependences since the functions $\hat\Sigma_F$ and $\hat\Sigma_\rho$ are to be involved in convolution integrals involving large values of, e.g.,  $\ln(p/p')$.} and where we have used ${\cal P}_\nu^{2\eta}(x)\approx{\cal P}_{\nu-2\eta}(x)={\cal P}_{\bar\nu-2\eps}(x)\approx{\cal P}_{\bar\nu}^{2\eps}(x)$ for $|x|\gtrsim1$ in obtaining the last expression. 
 
Equations \eqn{appeq:sigma_F} and \eqn{appeq:sigma_rho} are the results quoted in the text, Eqs.~\eqn{eq:sigmaF} and \eqn{eq:sigmarho}.

\section{Propagator in spacetime}
\label{appsec:real}

Here, we consider the spacetime structure of the NLO propagator obtained in the present work, Eqs.~\eqn{eq:Fmassive}--\eqn{eq:rho_massive}. The latter can be written as a linear superposition of two free massive propagators with masses $m_\pm\ll1$. We thus consider first the spacetime structure of the propagator of a free light scalar field in its CTBD vacuum. In general, this can be expressed exactly in terms of a hypergeometric function \cite{Chernikov:1968zm,Bunch:1978yq} but the case of a light field yields a much simpler form, both in the coincidence limit and at large spacetime separations, that we now discuss.

The statistical propagator of a scalar field with mass $m\ll1$ receives contributions $\sim 1/m^2$ which are generated by the infrared power law \eqn{eq:f_loc_ir}. Let us first consider the case of large spacelike separations, with spatial comoving distance $|{\bf X}-{\bf X}'|^2\gg \eta^2+\eta^{\prime2}$. In that case, the oscillating phase in \Eqn{eq:fourier} effectively cuts off the momentum integral at $K|{\bf X}-{\bf X}'|\lesssim 1$, and it is fully justified to use the leading infrared behavior of the correlator for $|K\eta|,|K\eta'|\ll1$, given by \Eqn{eq:f_loc_ir}. Using Eqs.~\eqn{eq:fourier}, \eqn{eq:prepG}, and \eqn{eq:f_loc_ir}, we have
\begin{align}
 F_m^{\rm IR}(x,x')&\approx (\eta\eta')^\varepsilon F_\nu\int\frac{d^d K}{(2\pi)^d}\frac{e^{i{\bf K}\cdot({\bf X}-{\bf X}')}}{K^{2\nu}}\nn
\label{appeq:IRspace}
 &\approx\frac{1}{\Omega_{D+1}m^2}\left(\frac{\eta\eta'}{|{\bf X}-{\bf X}'|^2}\right)^\varepsilon,
\end{align}
where $\nu=d/2-\eps$, with $\eps\approx m^2/d$, and where we used
\begin{align}
 \int\frac{d^dK}{(2\pi)^d}\frac{e^{i{\bf K}\cdot{\bf X}}}{K^{d-2\varepsilon}}&=\frac{1}{(4\pi)^{d/2}}\frac{\Gamma(\varepsilon)}{\Gamma(d/2-\varepsilon)}\left(\frac{2}{|{\bf X}|}\right)^{2\varepsilon}\\
 &\approx\frac{1}{(4\pi)^{d/2}\Gamma(d/2)}\frac{1}{\varepsilon |{\bf X}|^{2\varepsilon}}.
\end{align}

In the case of large timelike separations $|{\bf X}-{\bf X}'|^2\ll \eta^2+\eta^{\prime2}$, the phase factor $e^{i{\bf K}\cdot({\bf X}-{\bf X}')}\approx1$ in the relevant integration region and the integral in \eqn{eq:fourier} must effectively be cut off at $K\le\min\left(\frac{\mu}{|\eta|},\frac{\mu}{|\eta'|}\right)$ for the use of the leading infrared form \eqn{eq:f_loc_ir}---which yields the $1/m^2$ contribution to the correlator in spacetime---to be justified. For large time separation, i.e., $|t-t'|=|\ln(\eta/\eta')|\gg1$, we can replace the upper bound by $K^2\lesssim\frac{\mu^2}{\eta^2+\eta^{\prime2}}$, and we obtain
\beq
\label{appeq:IRtime}
 F_m^{\rm IR}(x,x')\approx\frac{1}{\Omega_{D+1}m^2}\left(\frac{\eta\eta'}{\eta^2+\eta^{\prime2}}\right)^\eps=\frac{1}{\Omega_{D+1}m^2}e^{-\eps|t-t'|},
\eeq
where we used
\beq
 \int\frac{d^dK}{(2\pi)^d}\frac{\theta(\Lambda-K)}{K^{d-2\varepsilon}}=\frac{\Omega_d}{(2\pi)^d}\frac{\Lambda^{2\eps}}{2\eps}=\frac{1}{(4\pi)^{d/2}\Gamma(d/2)}\frac{\Lambda^{2\eps}}{\eps}.
\eeq

Of course, the correlator is a function of the de Sitter invariant
\begin{align}
 z&=\frac{-\eta^2-\eta^{\prime2}+|{\bf X}-{\bf X}'|^2}{2\eta\eta'}\\
 &=-\cosh(t-t')+\frac{|{\bf X}-{\bf X}'|^2}{2}e^{t+t'}.
\end{align} 
The expressions \eqn{appeq:IRspace} and \eqn{appeq:IRtime} simply rewrite, up to relative corrections ${\cal O}(\eps)$,
\beq
\label{appeq:IRz}
 F_m^{IR}(x,x')\approx\frac{1}{\Omega_{D+1}m^2|z|^\varepsilon} \quad{\rm for}\quad |z|\gg1
\eeq
in the relevant limits of large spacelike or timelike separations. The leading $1/m^2$ behavior \eqn{appeq:IRz} can also be obtained directly from the expression of the exact propagator in terms of hypergeometric functions \cite{Garbrecht:2013coa}.

Finally, we note that the expression \eqn{appeq:IRz} also holds in the coincidence limit $z\to1$, for which
\beq
 F_m^{\rm IR}(x,x)\approx\eta^{2\eps}F_\nu\int\frac{d^d K}{(2\pi)^d}\frac{\theta(\mu-|K\eta|)}{K^{2\nu}}\approx\frac{1}{\Omega_{D+1}m^2}.
\eeq

Using the above considerations, we conclude that the NLO statistical propagator obtained in the present paper [\Eqn{eq:Fmassive}] behaves as
\beq
 F^{\rm IR}(x,x')\approx\frac{1}{\Omega_{D+1}}\left\{\frac{c_+}{m_+^2|z|^{\varepsilon_+}}+\frac{c_-}{m_-^2|z|^{\varepsilon_-}}\right\}
\eeq
for large spacetime separations $|z|\gg1$. In the coincidence limit we have
\beq
 F^{IR}(x,x)\approx\frac{1}{\Omega_{D+1}}\left\{\frac{c_+}{m_+^2}+\frac{c_-}{m_-^2}\right\},
\eeq
which reproduces the second line of \Eqn{eq:variance} up to relative corrections ${\cal O}(m_\pm^2)$.

\section{The field variance}
\label{appsec:variance}

This appendix is devoted to the calculation of the local field variance $\langle\varphi^2(x)\rangle/N$ at NLO in the $1/N$ expansion in both the stochastic and the Euclidean de Sitter 
approaches. 

\subsection{Stochastic approach}

In the stochastic approach, the long wavelength part of a light quantum field is treated as a slowly evolving classical stochastic field---owing to the strong gravitational enhancement of superhorizon modes---sourced by a white Gaussian random noise representing the quantum short wavelength (subhorizon) degrees of freedom which constantly cross out the horizon.\footnote{This approach neglects the self-interactions of subhorizon modes as well as the interaction between sub- and superhorizon modes, which is similar in spirit to the approximation strategy employed in the present work.} Neglecting spatial gradients, this is described by the following Langevin equation \cite{Starobinsky:1994bd,Beneke:2012kn}
\beq
\partial_t{\varphi_a} (t)+ \frac{1}{d}\partial_{\varphi_a} V(\varphi(t)) = \xi_a(t),
\label{eq:starobinsky}
\eeq
where $t=-\ln(-\eta)$ is the cosmological time, $V(\varphi)$ is the relevant potential for superhorizon modes \cite{Guilleux:2015pma}, and the Gaussian random noise is characterized by the correlator
\beq
\la \xi_a(t)\xi_b(t^\prime) \ra = \frac{2}{d \Omega_{D+1}}\delta_{ab}\delta(t-t^\prime).
\label{eq:noise_correlation}
\eeq
Here, the precise numerical prefactor depends on the (quantum) state of the subhorizon modes. The above (standard) expression assumes the CTBD vacuum.

The Langevin equation \eqn{eq:starobinsky} can be turned into a Focker-Planck equation for the probability distribution $\varrho(t,\varphi)$ of the stochastic process. The latter admits an attractor solution at late times, given by \cite{Starobinsky:1994bd}
\beq
\lim_{t\to\infty}\varrho(t,\varphi) \propto e^{-\Omega_{D+1} V(\varphi)}.
\eeq
The late-time expectation value of any local observable $\mathcal{O}(\varphi)$ is then given by 
\beq
\label{appeq:prescription}
 \la\mathcal{O}(\varphi) \ra = \frac{\int d^N\varphi \, \mathcal{O}(\varphi)\, e^{-\Omega_{D+1} V(\varphi)}}{\int d^N\varphi \,e^{-\Omega_{D+1} V(\varphi)}}.
\eeq 

In the case of interest here, where the potential is\footnote{Here, the square mass and coupling parameters are to be understood as the effective (renormalized) ones at the horizon scale $\mu\approx 1$ \cite{Guilleux:2015pma}.}
\beq
V(\varphi) = \frac{m_{\mathrm{dS}}^2}{2}\varphi_a\varphi_a +\frac{\lambda}{4!N}(\varphi_a\varphi_a)^2,
\eeq 
the field variance can be written as
\beq
 \la \varphi_a\varphi_b \ra = -\frac{\delta_{ab}}{N} \partial_{\alpha} \ln{\cal Z}(\alpha, \beta),
\eeq
where we introduced the notations $\alpha=\Omega_{D+1}m_{\rm dS}^2/2$ and $\beta=\Omega_{D+1}\lambda/24$ and where we defined, introducing the change of integration variable $\chi=\varphi^2/N$,
\beq
\label{appeq:integral}
 {\cal Z}(\alpha,\beta)=\frac{N^{N/2}}{2}\int_0^\infty d\chi\,e^{-Nf(\chi)},
\eeq
with
\beq
 f(\chi)=\alpha\chi+\beta\chi^2-\left(1-\frac{2}{N}\right)\frac{\ln\chi}{2}.
\eeq
The $1/N$ expansion of the integral \eqn{appeq:integral} is equivalent to a saddle-point expansion. The saddle point $\chi=\bar\chi$ is defined as $f'(\bar\chi)=0$, which is solved as 
\beq
 \bar\chi=-\frac{\alpha}{4\beta}+\sqrt{\left(\frac{\alpha}{4\beta}\right)^2+\frac{1}{4\beta}\left(1-\frac{2}{N}\right)}.
\eeq
The LO solution is, thus,
\beq
 \bar\chi_0=\frac{-\alpha+\sqrt{\alpha^2+4\beta}}{4\beta}=\frac{1}{\alpha+\sqrt{\alpha^2+4\beta}}
\eeq
and we shall only need the NLO correction
\beq
 \bar\chi=\bar\chi_0-\frac{1}{N}\frac{1}{\sqrt{\alpha^2+4\beta}}+{\cal O}\left(N^{-2}\right).
\eeq

The NLO result for the integral \eqn{appeq:integral} is given by the Gaussian integration around the saddle point,
\beq
 \frac{\ln{\cal Z}(\alpha,\beta)}{N}={\rm const}-f(\bar\chi)-\frac{1}{2N}\ln f''(\bar\chi_0)+{\cal O}\left(N^{-2}\right),
\eeq
from which it follows, using the definition $f'(\bar\chi)=0$, that
\beq
 -\frac{\partial_\alpha\ln{\cal Z}(\alpha,\beta)}{N}=\bar\chi+\frac{\partial_\alpha\bar\chi_0}{2N}\frac{f'''(\bar\chi_0)}{f''(\bar\chi_0)}+{\cal O}\left(N^{-2}\right).
\eeq
The calculation of each individual part is straightforward. We obtain
\begin{align}
 \frac{\la\varphi^2\ra}{N}&=\bar\chi_0\left(1-\frac{2}{N}\frac{\beta}{\alpha^2+4\beta}\right)\nn
 &=\frac{1}{\Omega_{D+1}M_0^2}\left(1-\frac{2}{N}\frac{\leff}{(1+\leff)^2}\right),
\end{align}
where we used the definitions \eqn{eq:LOmass} and \eqn{eq:effcoup} in obtaining the second line. This exactly agrees with the result of the present work from the solution of the KBEs, Eqs.~\eqn{eq:def} and \eqn{eq:mdyn0}.

\subsection{Euclidean de Sitter space}

For light scalar fields with nonderivative interactions, it has been shown in Refs.~\cite{Rajaraman:2010xd,Beneke:2012kn} that the effective theory of the field zero mode on the $D$-dimensional unit sphere $S_D$ (Euclidian de Sitter space) is equivalent to the stochastic prescription \eqn{appeq:prescription} for what concerns the calculation of the local field fluctuations. Still, it is instructive to compute explicitly the variance of the zero mode at NLO in the $1/N$ expansion directly in Euclidean de Sitter space. 

The line element is now given by (in $D$-dimensional spherical coordinates)
\beq
ds^2 = \mathrm{d}\theta^2 + \mathrm{sin}^2\theta\, \mathrm{d}\Omega_d.
\eeq
As in the main text, we consider the symmetric phase in which $\la\varphi_a\ra=0$ and where two-point functions are diagonal  in $O(N)$ space, e.g., $G_{ab}=\delta_{ab}G$. The Euclidean KBEs reads
\begin{align}
 \left(-\Box_x + m_{\rm dS}^2 \right)G(x,x') +\!\int_y \Sigma(x,y)G(y,x')= \delta^{(D)}(x,x'),
 \label{eq:SD_euclidian}
\end{align}
where $\Box_x$ is the appropriate Laplace-Beltrami operator, $\int_y=\int d^Dy\sqrt{g(y)}$, and $\delta^{(D)}(x,x')=\delta^{(D)}(x-x')/\sqrt{g(x)}$. 

The NLO self-energy is given by\footnote{Here, both the local and the nonlocal contributions involve the full propagator $G$. Strictly speaking this corresponds to the $1/N$ expansion at NLO in the two-particle-irreducible formalism \cite{Berges:2001fi,Aarts:2002dj}. The standard NLO contributions are obtained from the latter by systematically expanding the propagator around its LO expression, as we do below.}
\begin{align}
\Sigma(x,x^\prime) &=  \frac{\lambda(N+2)}{6N} G(x,x)\delta^{(D)}(x,x')\nn
&+\frac{\lambda}{3N}G(x,x^\prime)I(x,x^\prime),
\end{align}
where the function $I(x,x')$ resums the infinite series of bubble diagrams, as in \Fig{fig:NLO1}. It solves the integral equation 
\beq
I(x,x') = \Pi(x,x') + \int_y \Pi(x,y)I(y,x'),
\eeq
with the one-loop bubble
\beq
 \Pi(x,x')=-\frac{\lambda}{6}G^2(x,x').
\eeq

On the $D$-dimensional sphere, the field can be decomposed as
\beq
 \varphi_a(x)=\sum_{\vec L}\varphi_{a,\vec L}Y_{\vec L}(x),
\eeq
where $\vec L=(L,L_{D-1},\ldots,L_1)$ is a vector of integer numbers with $L\ge L_{D-1}\ge\cdots\ge|L_1|$ and where the $D$-dimensional spherical harmonics satisfy
\beq
 \square_x Y_{\vec L}(x)=-L(L+D-1)Y_{\vec L}(x)
\eeq 
and are normalized as
\beq
 \int_xY_{\vec L}^*(x)Y_{\vec L'}(x)=\delta_{\vec L,\vec L'}.
\eeq
The zero mode is the constant $Y_{\vec 0}=1/\sqrt{\Omega_{D+1}}$, with $\Omega_{D+1}$ the volume of the unit sphere $S_D$.

The variance of the zero mode of a free scalar field is $\propto 1/m_{\rm dS}^2$. Hence, the zero mode of light fields in units of the sphere radius undergo strong fluctuations and must be treated nonperturbatively. Following Refs.~\cite{Rajaraman:2010xd,Beneke:2012kn}, the dominant contributions to the corresponding effective theory is obtained by simply discarding all nonzero modes. The contributions of the latter can then be controlled by perturbative means. In particular, we write
\beq
 \varphi_a(x)=\bar\varphi_a+\check\varphi_a(x)\to\bar\varphi_a,
\eeq
with $\bar\varphi_a=\varphi_{a,\vec 0}Y_{\vec 0}=\int_x\varphi_a(x)/\Omega_{D+1}$. Accordingly, we only retain the constant contributions to the various two-point functions involved in the KBEs \eqn{eq:SD_euclidian}, that is,
\beq
 G(x,x')=\bar G+\check G(x,x')\to\bar G,
\eeq
and similarly for $\Sigma(x,x')$, $I(x,x')$, and $\Pi(x,x')$.

The previous NLO expression in the zero-mode sector are then
\beq
\bar\Sigma = \frac{\lambda}{6}\left(1+\frac{2}{N}\right) \frac{\bar G}{\Omega_{D+1}} + \frac{\lambda}{3N}\bar G\bar I,
\eeq
with
\beq
\bar I = \bar \Pi + \Omega_{D+1} \bar \Pi \bar I = \frac{\bar \Pi}{1-\Omega_{D+1}\bar \Pi},
\eeq
where $\bar \Pi = -\frac{\lambda}{6}\bar G^2$. We have, finally,
\beq
\label{appeq:selfzero}
\bar \Sigma =  \frac{\lambda}{6}\frac{\bar G}{\Omega_{D+1}}\left(1+\frac{2}{N}\frac{1}{1+ \frac{\lambda}{6}\Omega_{D+1}\bar G^2}\right).
\eeq

Similarly the KBEs \eqn{eq:SD_euclidian} in the zero-mode sector read
\beq
m_{\rm dS}^2\bar G+ \Omega_{D+1}\bar \Sigma\bar G= \frac{1}{\Omega_{D+1} }  ,
\eeq
which can be rewritten as
\beq
2\alpha \bar G +4\beta \bar G^2\left( 1 + \frac{2}{N}\frac{1}{1 +4\beta \bar G^2} \right) = 1
\eeq
using \Eqn{appeq:selfzero} and the definitions $\alpha=\Omega_{D+1}m_{\rm dS}^2/2$ and $\beta=\Omega_{D+1}\lambda/24$ introduced in the previous subsection.

This is easily solved at NLO in $1/N$,
\beq
\label{appeq:sol1}
\bar G = \frac{\la\bar\varphi^2\ra}{N} =\bar G_0\left(1- \frac{2}{N}\frac{\beta}{\alpha^2+ 4\beta}\right),
\eeq
with the LO solution
\beq
\label{appeq:sol2}
\bar G_0 = \frac{1}{\alpha + \sqrt{\alpha^2 + 4 \beta}}.
\eeq
As expected, Eqs.~\eqn{appeq:sol1} and \eqn{appeq:sol2} reproduce the stochastic result of the previous subsection and, therefore, the solution of the KBEs in Lorentzian
de Sitter spacetime obtained in the present paper.

\end{document}